\documentclass[aps,twocolumn,10pt,longbibliography,showpacs,amsmath,amssymb,superscriptaddress]{revtex4-1}

\usepackage{graphicx}
\usepackage{color}
\usepackage{subfigure}
\usepackage{dcolumn}
\usepackage{bm}
\usepackage{epsf}
\usepackage{amsmath,amstext,amssymb}
\usepackage{enumitem}
\usepackage{hyperref}
\usepackage[ansinew]{inputenc}
\usepackage{epstopdf}
\usepackage{appendix}
\usepackage{upgreek}
\newcommand{\be}{\begin{equation}}
\newcommand{\ee}{\end{equation}}
\newcommand{\bea}{\begin{eqnarray}}
\newcommand{\eea}{\end{eqnarray}}

\begin{document}

\title{ Catching, trapping and {\it in-situ}-identification of thorium ions inside Coulomb crystals of $^{40}$Ca$^+$ ions}

\author{Felix Stopp}
\email{felstopp@uni-mainz.de}
\affiliation{QUANTUM, Institut f\"ur Physik, Johannes Gutenberg-Universit\"at Mainz, 55128 Mainz, Germany}

\author{Karin Groot-Berning}
\affiliation{QUANTUM, Institut f\"ur Physik, Johannes Gutenberg-Universit\"at Mainz, 55128 Mainz, Germany}

\author{Georg Jacob}
\affiliation{Alpine Quantum Technologies GmbH, 6020 Innsbruck, Austria}
\affiliation{QUANTUM, Institut f\"ur Physik, Johannes Gutenberg-Universit\"at Mainz, 55128 Mainz, Germany}

\author{Dmitry Budker}
\affiliation{QUANTUM, Institut f\"ur Physik, Johannes Gutenberg-Universit\"at Mainz, 55128 Mainz, Germany}
\affiliation{Helmholtz Institut Mainz, 55099 Mainz, Germany}
\affiliation{Department of Physics, University of California, Berkeley, California 94720-7300, USA}
\affiliation{PRISMA Cluster of Excellence, Johannes Gutenberg-Universit\"at Mainz, 55128 Mainz, Germany}

\author{Raphael Haas}
\affiliation{Helmholtz Institut Mainz, 55099 Mainz, Germany}
\affiliation{Institut f\"ur Kernchemie, Johannes Gutenberg-Universit\"at Mainz, 55128 Mainz, Germany}

\author{Dennis Renisch}
\affiliation{Helmholtz Institut Mainz, 55099 Mainz, Germany}
\affiliation{Institut f\"ur Kernchemie, Johannes Gutenberg-Universit\"at Mainz, 55128 Mainz, Germany}

\author{J\"org Runke}
\affiliation{Institut f\"ur Kernchemie, Johannes Gutenberg-Universit\"at Mainz, 55128 Mainz, Germany}
\affiliation{GSI Helmholtzzentrum f\"ur Schwerionenforschung GmbH, 64291 Darmstadt, Germany}

\author{Petra Th\"orle-Pospiech}
\affiliation{Helmholtz Institut Mainz, 55099 Mainz, Germany}
\affiliation{Institut f\"ur Kernchemie, Johannes Gutenberg-Universit\"at Mainz, 55128 Mainz, Germany}

\author{Christoph E. D\"ullmann}
\affiliation{Helmholtz Institut Mainz, 55099 Mainz, Germany}
\affiliation{PRISMA Cluster of Excellence, Johannes Gutenberg-Universit\"at Mainz, 55128 Mainz, Germany}
\affiliation{Institut f\"ur Kernchemie, Johannes Gutenberg-Universit\"at Mainz, 55128 Mainz, Germany}
\affiliation{GSI Helmholtzzentrum f\"ur Schwerionenforschung GmbH, 64291 Darmstadt, Germany}

\author{Ferdinand Schmidt-Kaler}
\affiliation{QUANTUM, Institut f\"ur Physik, Johannes Gutenberg-Universit\"at Mainz, 55128 Mainz, Germany}
\affiliation{Helmholtz Institut Mainz, 55099 Mainz, Germany}
\affiliation{PRISMA Cluster of Excellence, Johannes Gutenberg-Universit\"at Mainz, 55128 Mainz, Germany}

\begin{abstract}
Thorium ions exhibit unique nuclear properties with high relevance for testing symmetries of nature, and Paul traps feature an ideal experimental platform for performing high precision quantum logic spectroscopy. Loading of stable or long-lived isotopes is well-established and relies on ionization from an atomic beam. A different approach allows trapping short-lived isotopes available as alpha-decay daughters, which recoil from a thin sample of the precursor nuclide. A prominent example is the short-lived $^{229\text{m}}$Th, populated in a decay of long-lived $^{233}$U. Here, ions are provided by an external source and are decelerated to be available for trapping. Such setups offer the option to trap various isotopes and charge states of thorium. Investigating this complex procedure, we demonstrate the observation of single $^{232}$Th$^+$ ions trapped, embedded into and sympathetically cooled via Coulomb interactions by co-trapped $^{40}$Ca$^+$ ions. Furthermore, we discuss different options for a non-destructive identification of the sympathetically cooled thorium ions in the trap, and describe in detail our chosen experimental method, identifying mass and charge of thorium ions from the positions of calcium ions, as their fluorescence is imaged on a CCD camera. These findings are verified by means of a time-of-flight signal when extracting ions of different mass-to-charge ratio from the Paul trap and steering them into a  detector.  
\end{abstract}


\maketitle

\section{Introduction}
\label{intro}

High precision spectroscopy allows for tests of fundamental symmetries of nature. In the past such investigations have been based on ultra-high precision measurements in elementary atomic systems, such as single electrons, protons, or hydrogen atoms. The accuracy of modern frequency standards allows for tests at a level of a few parts in 10$^{18}$. The modern techniques of trapping and cooling have been essential to achieve such results. Exotic atomic systems such as highly charged ions, atomic systems with exotic nuclear properties, or systems with antimatter particles will complement existing tests with elementary simple atomic systems. However, there are several additional challenges: i) Exotic atomic systems are not available in all labs, and typically they come in very small quantities, and/or with low production rates. ii) They require a complex fabrication process: typically one has a separate source apparatus for their fabrication and faces thus the challenge to transport and catch such exotic systems in the trapping region for spectroscopic investigations. iii) In many cases, sympathetic cooling is required to prepare them for spectroscopy. iv) Identifying the successful trapping and cooling of such atomic systems may need additional auxiliary atoms or ions, and if v) a direct observation may be hard or impossible to implement, quantum logic spectroscopy is required~\cite{Wineland}. 
Here we describe initial steps towards such research program for the specific case of trapped thorium ions. This element features a number of different isotopes, including $^{229}$Th, which features a uniquely low-lying nuclear isomer at an excitation energy on the eV scale, accessible for optical techniques ~\cite{Wense,Seiferle}. We describe how to catch single thorium ions from an external source into a Paul trap, observe and identify the thorium in-situ and in non-destructive manner. Recoil sources for thorium ions, currently under development, will be able to deliver them in a large variety of charge states~\cite{Gunter}. We first sketch the experimental setup and the capture of thorium ions, and then discuss the options for detection. In the outlook, we briefly discuss next steps of the experiment.

\section{Experimental setup}
\label{exp}

\begin{figure*}
\includegraphics[width=0.7\textwidth]{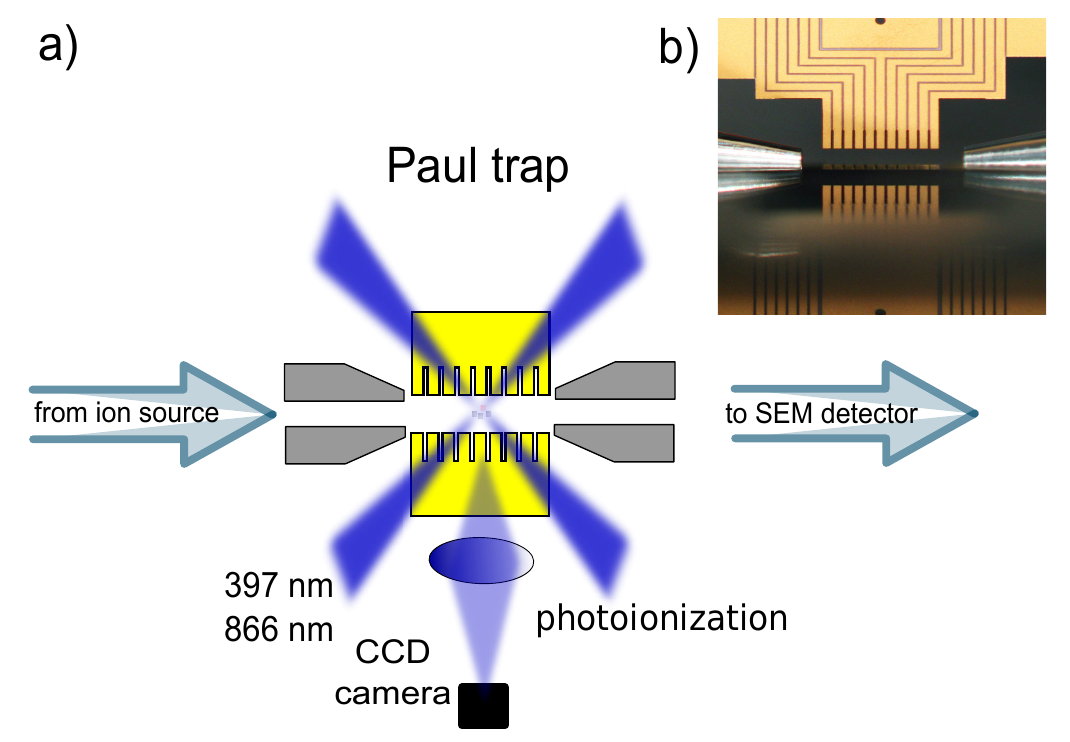}
\caption{a) Sketch of the experimental setup, which includes an ion source for creation, a linear Paul trap, and a SEM detector for detection of single thorium ions. b) Photo of the two endcaps (grey) with a distance of 2.9 mm and the blades with eleven DC segments (yellow). Each segment has a width of $200\,\upmu\text{m}$, the radial distance of the blades is $590\,\upmu\text{m}$. }
\label{setup}       
\end{figure*}

The experimental setup consists of three parts. (i) In the source, neutral thorium atoms are ablated by laser pulses from a metal target, ionized, subsequently accelerated and guided to the Paul trap. (ii) Ions are captured in a Paul trap, see Fig.~\ref{setup}. To achieve this, we apply optimized deceleration voltage ramps to the entrance endcap of the trap \cite{Groot}. Ions are then confined in the dynamic and static potential of the segmented linear Paul trap, see Fig.~1. The trap features two chips  each with eleven DC segments  (see Fig. \ref{setup}.b), yellow) and two RF electrodes identically shaped as the the DC electrodes and thus forming a X-blade design. Endcaps (grey)  at a distance of $2.9\,\text{mm}$ on both sides of the trap provide axial confinement. The RF electrodes operate with a radio frequency of $\Omega_\text{RF}= 2\pi \cdot 23.062\,\text{MHz}$ and a peak to peak amplitude $U_\text{pp}\approx570\,\text{V}$ to stabilize ion crystals inside of the trap. The trapping parameters are optimized for $^{40}$Ca$^+$ ion crystals. We load the trap using a neutral calcium atom beam and photoionize via two-step excitation near 423~nm and 375~nm  \cite{Jacob}. For continuous laser cooling of calcium ions, we use the  S$_{1/2}$ to P$_{1/2}$ transition and tune the 397~nm laser about 20~MHz to the red of the transition frequency. Laser radiation near 866~nm is used to empty the metastable D$_{3/5}$ level and we obtain continuous fluorescence which is imaged on an electron multiplying charge-coupled device (EMCCD) camera. In this crystal the thorium ions which have been injected from the external source are loaded and sympathetically cooled. We do not apply laser radiation on transitions of this element. Consequently, these ions appear as dark voids in the fluorescence image. (iii) When applying  a voltage ramp to the exit endcap of the trap, ions may be extracted, and travel about 30~cm distance before impacting into a a secondary electron multiplier (SEM) detector which registers the impact. Via time-of-flight measurement the ion mass-to-charge ratio is determined~\cite{Groot}. Even though this method provides an accurate identification at the single ion level, it is destructive, and may only be used as a cross-check for verifying the previous non-destructive observation and identification of single thorium ions. The experimental challenges are of similar sort as compared to trapping highly charged ions for spectroscopy that are generated in an EBIT~\cite{Schmoger,Schmoger2}. 

\section{Detection and identification of co-trapped ions}
\label{sim}
In the following we discuss different methods of non-destructive detection: \\
(i) For sensing the {\bf frequencies of common modes} of a mixed two-ion crystal and deducing from this the mass of the co-trapped and sympathetically cooled ion one may apply sideband spectroscopy on a narrow optical transition using a frequency stabilized laser~\cite{Leibfried,Welzel}. In this case the sub-kHz resolution is reached and allows for an unambiguous identification of mode frequencies and the masses. An alternative method is to excite one of the common modes with an oscillating voltage applied to one of the segmented electrodes and scan that frequency. Near resonance, the EMCCD image of the calcium ion will smear out. The latter method comes with the severe problem that small motional oscillations are hard to detect with the imaging system that features limited optical resolution of a few $\upmu$m. Exciting with higher electric amplitude leads to a nonlinear resonance shape, as a result of the balance between continuous Doppler cooling and electric motional excitation. Under such conditions, the determination of the resonance frequency is prone to systemic errors. A refined method relies on resonant switching of optical laser beam for Doppler cooling~\cite{Drewsen} and the experimental results demonstrate clearly sub-kHz resolution and a clear identification of the masses of the ions. \\
(ii) A completely different approach relies on the {\bf observation of structural changes}, which have been coined mass-defects~\cite{Ulm} and lead to a zigzag- or buckled shape of the ion crystal. As the position of the fluorescent calcium ions is altered by the presence of one or more co-trapped dark ions, comparison of measured and calculated positions allows  deducing the mass-to-charge ratio and the charge state of those species. We have concentrated on this method. An advantage of this method is that no additional lasers, especially no highly stabilized laser sources are required, and that this method can be easily implemented and is fast, as compared to scanning over sideband frequencies.

\section{Calculation of mixed-ion-crystal frequencies and positions}
\label{theo}

Mode frequencies of ion crystals with two different species have been calculated~\cite{Kielpinski,Morigi}. We introduce the mass ratio of two species $\mu_n = m_\text{Ca} / m_n$.
The two modes of the axial direction read ~\cite{Wubbena}:
\begin{align}
\omega_\text{ax-com./breath.}&=\sqrt{\frac{1+\mu_n\pm\sqrt{1-\mu_n+\mu_n^2}}{\mu_n}}\omega_\text{ax}.
\end{align}
Here, $\omega_\text{ax}$ is the axial trap frequency for one $^{40}$Ca$^+$ ion. Table 1 shows the calculation in the case of one $^{40}$Ca$^+$ ion trapped together with one ion of large mass. We have taken typical trapping parameters into account. Following the method (i), i.e., using sideband spectroscopy, distinguishing between ions of different mass, e.g. for identifying different isotopes, either the x- or the y-rocking modes are suited best, since these radial modes vary by about 2~kHz when changing the mass of the heavy ion by one atomic unit, u.

\begin{table}
\caption{Calculated mode frequencies of a two-ion crystal. The common and rocking/breathing mode values for a two $^{40}$Ca$^+$ ion crystal and one $^{40}$Ca$^+$ ion together with a second ion of mass 231, 232 or 233.}
\label{table}
 \begin{tabular}{l|c|c|c|c}
\hline\noalign{\smallskip}
$^{40}$Ca$^+$  +...&   $^{40}$Ca$^+$  & $^{231}$X$^+$ & $^{232}$X$^+$ & $^{233}$X$^+$   \\
\noalign{\smallskip}\hline\noalign{\smallskip}
 x-com.  & 1640 kHz & 1624 kHz & 1624 kHz & 1624 kHz\\
 y-com.  & 1758 kHz & 1744 kHz & 1744 kHz & 1744 kHz\\
 ax-com. & 320 kHz & 159.2 kHz & 158.9 kHz & 158.5 kHz\\
 x-rocking &  1608.4 kHz & 188.1 kHz & 186.6 kHz & 185.2 kHz\\
 y-rocking & 1728.6 kHz & 324.4 kHz &323.1 kHz & 321.8 kHz\\
 ax-breathing & 554.3 kHz & 463.6 kHz & 463.5 kHz & 463.5 kHz\\
\noalign{\smallskip}\hline
 \end{tabular}
\end{table}

\begin{figure*}
\includegraphics[width=0.7\textwidth]{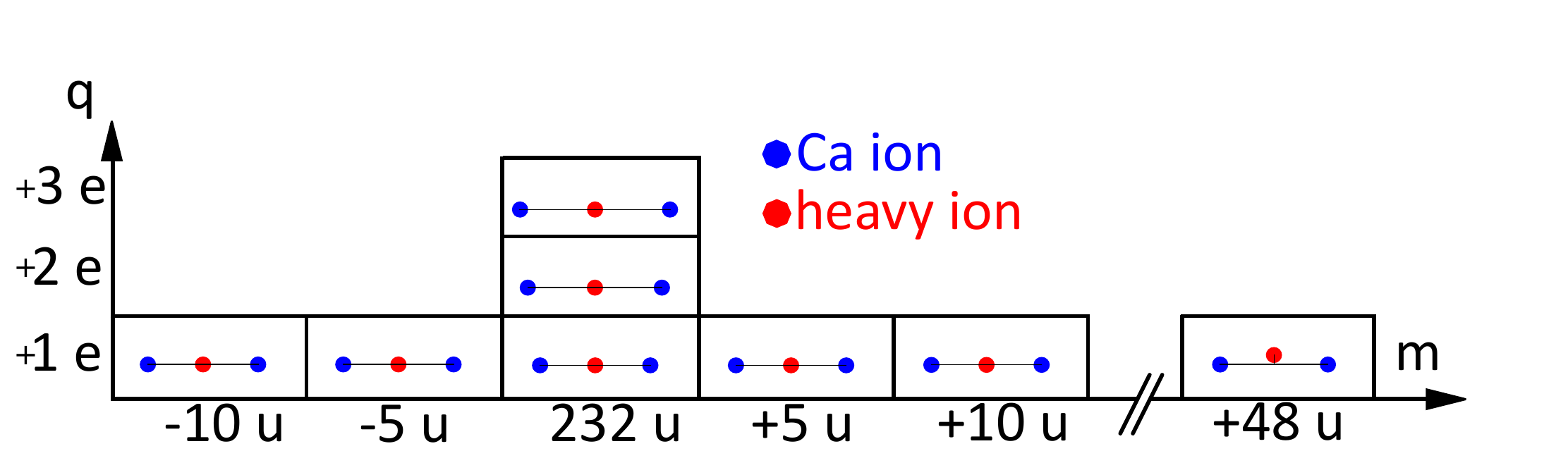}
\caption{Simulation of the mixed ion crystal positions, including two calcium ions (blue) with mass $40\, \text{u}$ and charge state  $+1\cdot e$, and one heavier ion (red) with different masses $m$ and charge states $q$.}
\label{three_ion}       
\end{figure*}

\begin{figure*}
\includegraphics[width=0.7\textwidth]{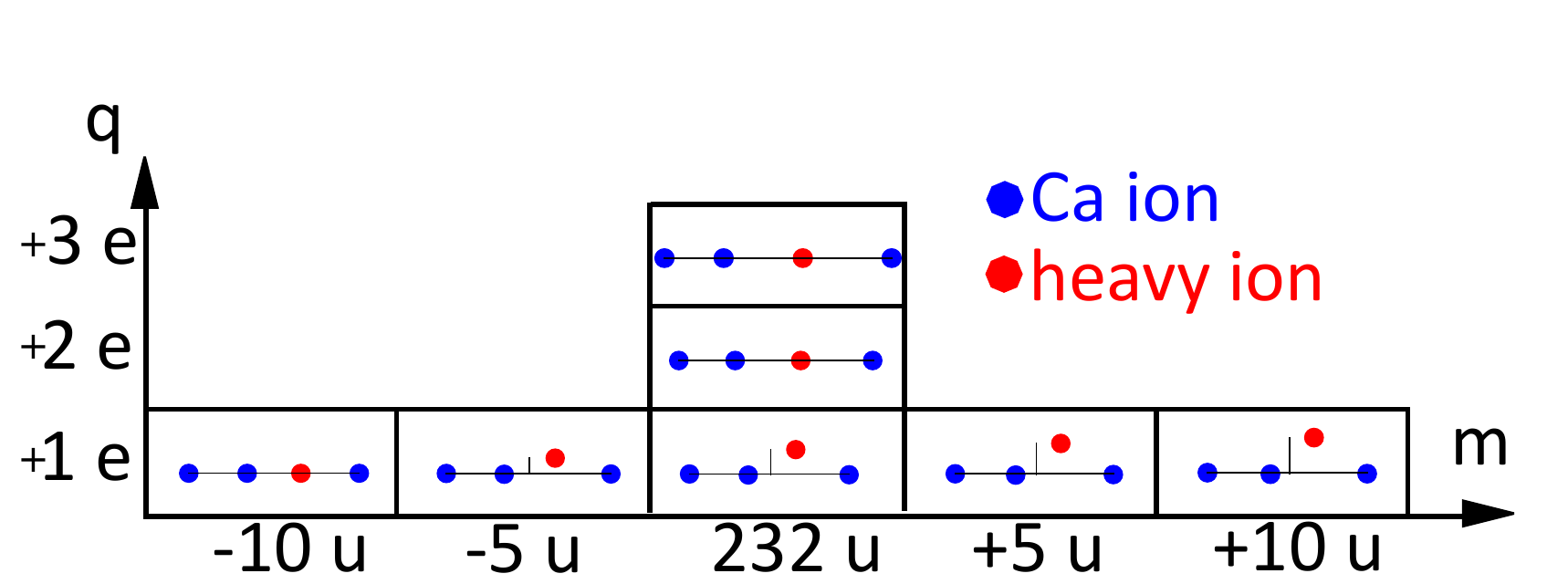}
\caption{Simulation of the mixed ion crystal positions, including three calcium ions (blue) with mass $40\,\text{ u}$ and charge state  $+1\cdot e$, and one heavier ion (red) with different masses $m$ and charge states $q$.}
\label{four_ion}       
\end{figure*}

When we follow the approach of observing structural changes in the EMCCD image, we first have to calculate how the masses $m_n$ of $N$ charged particles will arrange as a mixed crystal in a linear Paul trap,  and determine their equilibrium positions. The method is described in \cite{James,Home,Marquet}. The potential energy ist hereby given by the pseudo potential approximation and the Coulomb interaction between the different ions,
\begin{align}
U(\vec{x}_1...\vec{x}_N)=\frac{1}{2}\sum_{i=1}^{3}\sum_{n=1}^{N}m_n\omega^2_{i,n}x^2_{i,n}+....\nonumber\\
...\frac{Z_1 Z_2e^2}{8\pi\epsilon_0}\sum\limits_{\substack{n,l=1\\ n\neq l}}^N\frac{1}{\sqrt{\sum_{i=1}^{3}(x_{i,n}-x_{i,l})^2}}.
\end{align}
The trap frequencies $\omega_{i,n}$ for different ion species are given by
\begin{align}
\omega_{i,n}=\frac{\Omega_\text{RF}}{2}\sqrt{\frac{(\mu_n\, q_{i,0})^2}{2}+\mu_n a_{i,0}}
\end{align}
with the mass of the calcium ion $m_\text{Ca}$ and the trap parameters $q_{i,0}$ for the dynamical and $a_{i,0}$ for the static part with the condition $\sum_{i=1}^3a_{i,0}=0$, $q_{1,0}+q_{2,0}=0$ and $q_{3,0}=0$. Assuming an idealized linear Paul trap potential with no dynamical confinement along axial direction we follow this approximation here. The equilibrium positions $\bar{x}_{i,n}$ can be calculated by canceling the forces, which means to solve the equation system:
\begin{align}
\left[ \frac{\partial U}{\partial x_{i,n}}\right]_{x_{i,n}=\bar{x}_{i,n}}=0 \hspace{0.5cm} \forall\, i\in\{1,2,3\},\, n\in\{1...N\}
\end{align}
We solve this set of equations in Mathematica. Results of these calculations are shown in Fig.~\ref{three_ion} and \ref{four_ion}, where we used the trap frequencies of $\omega_\text{x-,y-,ax-com.}=2\pi\cdot (1.640,\, 1.758,\,0.320)\,\text{MHz}$ for calcium.
Assuming the impurity as a thorium ion with mass $232\,\text{u}$, the calculations demonstrate that the positions of the three ion crystal with charge state $+1\cdot e$ would not change into a buckled configuration until it reaches a mass of $280\,\text{u}$, which accordingly is an upper limit for the mass. The distances do not change noticeably around the mass $232\,\text{u}$. Higher charge states also remain in a linear symmetric configuration, the distance between the two outer $^{40}$Ca$^+$ ions increases with growing charge state.\\
A mixed four ion crystal including mass $232\,\text{u}$ already skips into a buckled configuration for this trap anisotropy of $\alpha=\omega_\text{ax}/\omega_\text{rad} =0.19$ and the phase transition takes place at mass values of around $227\,\text{u}$, see Fig.~\ref{four_ion}. For an uncertainty of $10\,\text{kHz}$ in determining all three trap frequencies, we can define a lower limit of the mass with $m\geq230(5)\,\text{u}$. However, higher charge states remain in a linear configuration, but the symmetry of the distances to each other is broken, due to the even number of ions, but odd number of impurities.\\
Thus, we can determine a limitation of the mass based on the phase transition between linear and buckled configuration and thus identify the impurity as heavy ion, which stands out clearly from background gases, as long as it has the charge state $+1\cdot e$. The exact mass determination is limited by the fluctuations in the trap frequencies, since even a change if $10\,\text{kHz}$ in axial direction means a fluctuation of $5\, \text{u}$.\\
Mixed crystals with higher charge states in linear configuration can be precisely determined. Nor are they limited by the imaging system (see section \ref{obs}) because for any higher charge state the distance of the outer ions increase by more than two $\upmu$m.


\section{Observation and experimental analysis of pure and mixed ion crystals}
\label{obs}

\begin{figure*}
\centering
\includegraphics[width=0.4\textwidth]{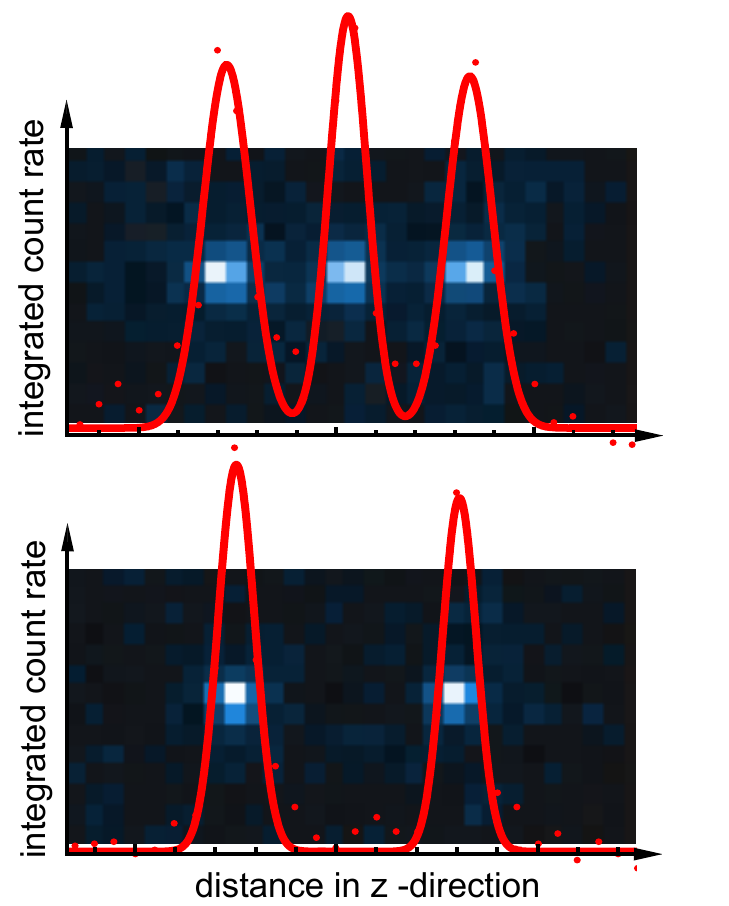}
\caption{Comparison of a three ion crystal consisting of pure $^{40}$Ca$^+$ and including one $^{232}$Th$^+$. Both pure and mixed crystal form a linear configuration. Gaussian fits are used to determine z-positions. For this, the EMCCD camera was exposed $40\,\upmu\text{s}$ with the fluorescent scattered light and the total number of counts projected onto the z-axis. }
\label{three_comp}       
\end{figure*}

\begin{figure*}
\centering
\includegraphics[width=0.4\textwidth]{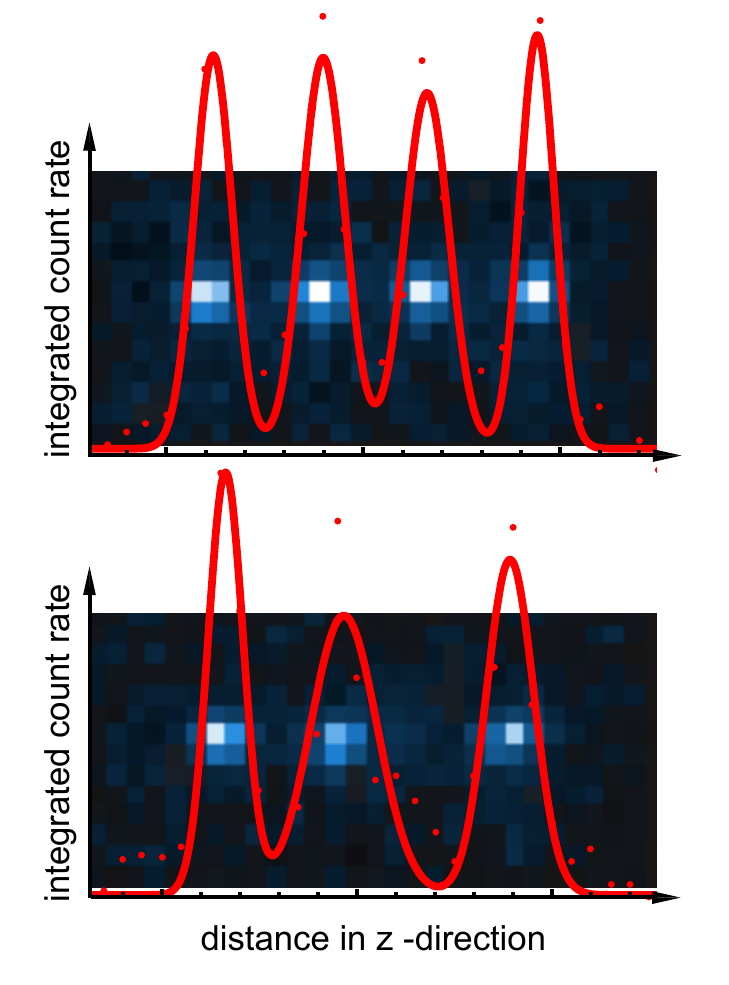}
\caption{Comparison of a four ion crystal consisting of pure $^{40}$Ca$^+$, top, and a crystal including one $^{232}$Th$^+$, bottom. The thorium ion forms a void spot in the crystal, not emitting fluorescence. The mixed crystal shows a buckled formation as calcium ions do not sit on a line. The second ion is shifted below, as the dark thorium ion is sitting above. Gaussian fits are used to determine z-positions.}
\label{four_comp}       
\end{figure*}

Initially, we determine the magnification of the imaging optics, for which we use a crystal of two calcium ions, imaged on the EMCCD camera. The size of one pixel of the EMCCD camera is  $24\,\upmu$m. Summing up the fluorescence data over the radial direction and fitting Gaussian functions, we determine a distance of $7.22(7)\, \text{px}$. Also, we determine the trap frequency $\omega_{z}= 320(10)~\text{kHz}$. The error is dominated by a small dynamical confinement in axial direction. Based on the calculation of equilibrium distances, see Sect.~\ref{theo}, the distance for a two-ion crystal with this axial frequency is $11.98\,\upmu\text{m}$. Thus, we determine an accounting factor of $1.66(2)\, \upmu\text{m}/\text{px}$ and a magnification of $14.5(5)$.\\
Figures \ref{three_comp} and \ref{four_comp} each show the comparison of a three and four ion crystal with and without impurity. Gaussian functions were added to the pixel data of the EMCCD camera. Therefor, the camera had an exposure time of $40\,\upmu\text{s}$, in which the scattered light hit the sensor. Subsequently, the brightness values were summed over the columns of the pixel matrix and projected onto the z-axis. The different heights  of the Gaussian functions can be explained by the different average brightness values for each ion but reflect the position of the ions better than one pixel size.\\
Under the operating conditions of the measured trap frequencies (see section \ref{theo}), which corresponds to an anisotropy of $\alpha=\omega_\text{ax}/\omega_\text{rad} =0.19$, a calcium ion crystal with up to $n_\text{Ca}=5$ ions still shows a linear shape. However, the experimental finding for a mixed crystal, including also one thorium and that for $n_\text{Ca}=3$, see Fig.~\ref{four_comp}, shows that this linear symmetry is broken and is passing over in a buckled configuration. \\
Thus, as any mass below $230(5)\,\text{u}$ is excluded, the dark impurity sites have to be $^{232}$Th$^+$. Due to the purity of the sample and the vacuum, no further elements with such a high mass number can be present in the apparatus. \\
Impurities coming from background gases and chemical reactions with $\text{Ca}^+$, e.g. hydrogen or oxygen is principally possible, but their highest mass would be $56\,\text{u}$ for CaO and is therefore low to be forced in a zigzag configuration by the anisotropy of the trap. The same applies to the impurities of the metal sample.\\
The charge state has to be $+1\cdot e$. Otherwise the configuration would also stay in a linear form, which figure \ref{four_ion} shows for mass $232\,\text{u}$ and charge state $+2\cdot e$ respectively $+3\cdot e$. These charge states may well be possible if one gains different thorium isotopes directly from the parent nucleus \cite{Gunter} and can be crystallized in a trap \cite{Schmoger}. In case of $n_\text{crystal}=3$ crystal, the anisotropy between radial and axial trap frequency is not strong enough and keeps the mixed crystal in a linear structure (see figure \ref{three_comp}).\\
The in-situ-identification was verified via a TOF measurement. Here, a voltage ramp at the endcaps steers the ion crystal to the SEM detector and determines the time of flight. A detailed model description of the determination of the masses from the flight times is described in \cite{Groot}. The results show a clear agreement.

\section{Conclusion}

The report briefly outlines two options of identifying a thorium ion in a mixed two-ion crystal in a non-destructive way and represents an extension of \cite{Groot}. Both the sideband spectroscopic approach and the in situ position determination provide a rapid response if a heavy ion is trapped. In the most cases this is perfectly adequate when conducting experiments with a particular isotope of thorium. The impurities in the $^{232}$Th$^+$ sample, with which these experiments were carried out, are only limited to Sn \cite{Groot}.\\
The advantages of a in-situ-identification with the pictures of the CCD camera should be mentioned again. Due to the fast exposure time, it is a fast process and can be evaluated directly and with a relatively high resolution. Also, this method can be implemented without much effort and requires no further devices, such as highly stabilized lasers as in the sideband spectroscopy.\\
This in turn is of high importance when looking in the direction of quantum logic spectroscopy. In conclusion, reference should be made to the hyperfine spectroscopy, e.g., to identify $^{229\text{m}}$Th with respect to $^{229}$Th \cite{Thielking}. It is planned to improve the accuracy of these measurements in mixed crystals and by using spectroscopy on individual cold thorium ions.\\
This work was financially supported by the Helmholtz Excellence Network ExNet-020, Precision Physics, Fundamental Interactions and Structure of Matter
(PRISMA+) from the Helmholtz Initiative and Networking Fund, and we acknowledge financial support by the DFG DIP program (FO 703/2-1) and by the VW Stiftung.

\end{document}